\documentclass[12pt,a4paper]{article}
\usepackage{graphicx}

\newcommand{\be}{\begin{equation}}
\newcommand{\ee}{\end{equation}}
\newcommand{\ba}{\begin{eqnarray}}
\newcommand{\ea}{\end{eqnarray}}
\newcommand{\rmi}[1]{{\mbox{\scriptsize #1}}}

\newcommand{\psat}{p_{\rmi{sat}}}

\newcommand{\msbar}{\overline{\mbox{\rm MS}}}
\newcommand{\agev}{\,A{\rm GeV}}

\def\lsim{\raise0.3ex\hbox{$<$\kern-0.75em\raise-1.1ex\hbox{$\sim$}}}
\def\gsim{\raise0.3ex\hbox{$>$\kern-0.75em\raise-1.1ex\hbox{$\sim$}}}

\begin{document}

\begin{titlepage}
\begin{flushright}
JYFL-5/00\\
hep-ph/0010319 \\
20 October, 2000\\
\end{flushright}
\begin{centering}
\vfill

{\bf TRANSVERSE ENERGY FROM MINIJETS \\ IN ULTRARELATIVISTIC
NUCLEAR COLLISIONS:\\ A NEXT-TO-LEADING ORDER ANALYSIS \\}
 
\vspace{1cm}
 K.J. Eskola\footnote{kari.eskola@phys.jyu.fi}$^{a,b}$
and K. Tuominen\footnote{kimmo.tuominen@phys.jyu.fi}$^{a}$ 

\vspace{1cm}
{\em $^a$ Department of Physics, University of Jyv\"askyl\"a, \\
P.O.Box 35, FIN-40351, Jyv\"askyl\"a, Finland \vspace{0.5cm}\\}
{\em $^b$ Helsinki Institute of Physics, \\
P.O.Box 9, FIN-00014, University of Helsinki, Finland\\} 
\vspace{1cm}
\end{centering}

\centerline{{\bf Abstract}}

\noindent We compute in next-to-leading order (NLO) perturbative QCD
the amount of transverse energy produced into a rapidity region
$\Delta Y$ of a nuclear collision from partons created in
the few-GeV subcollisions. The NLO formulation assumes collinear
factorization and is based on the subtraction method. We first study
the results as a function of the minimum transverse momentum scale and
define and determine the associated $K$-factors. The dependence of the
NLO results on the scale choice and on the size of $\Delta Y$ is also
studied. The calculations are performed for GRV94 and CTEQ5 sets of
parton distributions. Also the effect of nuclear
shadowing to the NLO results is investigated. The main results are
that the NLO results are stable relative to the leading-order (LO)
ones even in the few-GeV domain and that there exists a new
kinematical region not present in the LO, which is responsible for the
magnitude of the $K$-factors.  The dependence on the size of $\Delta
Y$ is found to be practically linear within two central units of
rapidity, and the scale dependence present at RHIC energies is seen to
vanish at the LHC energies. Finally, the effect of NLO corrections to
the determination of transverse energies and multiplicities in nuclear
collisions is discussed within the framework of parton saturation.

\vfill
\end{titlepage}


\section{Introduction} 

The heavy ion community experiences currently a new exciting period as
the first data in ultrarelativistic heavy ion collisions at BNL-RHIC
are being analysed and the first physics results are coming out. These
include the observation by the PHOBOS collaboration of an increased
activity in Au+Au collisions at $\sqrt s=56$ and $130 \agev$
\cite{PHOBOS}: the number of charged particles per participant pair
clearly increases from that in Pb+Pb collisions at the CERN-SPS at
$\sqrt s=17.3\agev$, and from $p\bar p$ at $\sqrt s=50\dots 200$
GeV. This lends support to a description of initial particle
production in high energy heavy ion collisions in terms of binary
collision mechanisms. The first results from the STAR collaboration
have in turn shown an increase of elliptic flow in Au+Au at $\sqrt
s=130\agev$, observed as an enhanced asymmetry in the azimuthal
distributions of particles \cite{STAR}. This points towards
collectivity, pressure and thermalization in the produced system.

A complete description of a collision of two heavy ions at high energy
is clearly an extremely challenging task: elementary QCD-quanta,
gluons, quarks and antiquarks, are expected to be produced abundantly
enough, so that they thermalize locally and form a new,
deconfined phase of matter, the Quark Gluon Plasma (QGP). After its
production the system will experience an expansion stage during which
it will cool down and go through a phase transition from the QGP to a
hadron gas. Eventually, decoupling sets in and the hadrons fly
independently to the detector, possibly decaying into other particles
before that.

Due to our ignorance of how to solve nonperturbative QCD, calculations
to describe such a complicated spacetime evolution are very difficult
to do from truly first principles. Instead, the evolution can be
effectively modelled in terms of relativistic hydrodynamics
\cite{BJORKEN,KATAJA}, which relies on the assumption of a locally
thermal system. The benefit in such an approach, in addition to
simplicity, is a proper treatment of the phase transition and
inclusion of collective effects (pressure and flow). The initial
conditions, however, have to be given from outside. Alternatively, one
may try to describe the strongly interacting system first as a cascade
of partons \cite{GEIGER,BIN} and then as a hadronic cascade
\cite{URQMD,BASS}, or, as recently suggested, as a combination of the
hydrodynamical approach and a hadron cascade \cite{DUMITRU1}.

Common to the approaches mentioned above, the initial conditions,
i.e. the energy densities and number densities of initially produced
quarks and gluons, along with the formation times, need to be known as
precisely as possible. As also the degree of thermalization of the QGP
\cite{DUMITRU2, MUELLER1, MUELLER2} depends heavily on the initial
conditions, it is vitally important to study them in detail.  In this
paper we will do this by formulating and computing the initial
transverse energy production from minijets in central $AA$ collisions
to next-to-leading order (NLO) in perturbative QCD (pQCD).

Initial particle production in $AA$ collisions at collider energies
has typically been described with two-component models where pQCD
accounts for the hard and semihard scatterings of partons and a
phenomenological model for the soft particle production
\cite{EKL89,XNW91,HIJING,FRITIOF}. At sufficiently high cms-energies,
partons produced with transverse momenta in the semihard domain,
$p_T\ge p_0\sim 1\dots 2$ GeV, become the dominant source of energy
deposited in central rapidities \cite{EKL89,KLL87,BM87}.  Within the
leading twist framework the inclusive cross sections for production of
semihard partons, along with the initial $E_T$ carried by them, are
computable in terms of collinear factorization \cite{CSS85} of pQCD
\cite{BM87,EKL89}. Individual binary scatterings of partons (or
nucleons) are assumed to be independent of each other and nuclear
effects can be included by using nuclear parton distributions
\cite{EKS98}.  Computation of the pQCD component thus leans on the
fact that production of quarks and gluons is reliably described by
pQCD at large transverse momentum scales, where it is possible to
actually observe individual jets in $pp$ collisions.  The main
uncertainties are related to the determination of the semihard scale
$p_0$ defining the perturbative component, and to the higher order
contributions to the parton cross sections. To reduce the latter
source of uncertainty is the main motivation for the present NLO
study.

The studies \cite{EKL89,XNW91,HIJING,KLL87} are in practise extensions
from the $pp\,(p\bar p)$ physics in the sense that the smallest
transverse momentum included in the perturbative component, $p_0$, is
constrained by data from hadronic physics and kept constant in the
extrapolation to $AA$. Recently, however, ideas of parton saturation
\cite{BM87,GLR} have been revived for an effective estimation of the
initial gluon and quark production based on the pQCD component
\cite{EKRT99,EKT00} alone.  The perturbative production of partons is
extendend down to a minimum scale $p_0=p_{\rm sat}$, defined as the
scale at which the produced partons, each of which is of the size
$\pi/p_0^2$, effectively fill the available transverse area $\pi
R_A^2$. This is the simplest way of defining a saturation criterion in
central $AA$ collisions. For $A\sim200$, this procedure leads to
$p_{\rm sat}\sim 1$ GeV for RHIC energies $\sqrt s= 56\dots 200\agev$
and $p_{\rm sat}\sim 2$ GeV for the LHC energy $\sqrt s = 5500\agev$
\cite{EKRT99}. The criterion which determines when the system becomes
saturated, may contain nonperturbative, even phenomenological
features. Terminating the pQCD computation at $p_0=p_{\rm sat}(\sqrt
s,A)$ thus gives an estimate of the total initial parton
production. The effective formation time of the minijet (gluon) plasma
system is then obtained as $\tau_0\sim 1/p_{\rm sat}$.  The
determination of $p_{\rm sat}$ in the pQCD+saturation approach is a
dynamical procedure which is affected by the NLO contributions to
partonic cross sections. Again, reducing the uncertainty related to
the higher order contributions in minijet production is the general
goal in this paper.

Another promising approach for describing initial gluon production is
that of classical fields \cite{McLV94}. The gluon fields originate
from colour sources moving along the light cones and, assuming
longitudinal boost-invariance, an effective dimensionally reduced
Hamiltonian can be formed and the gluon production computed through a
lattice simulation \cite{KV1}. Interestingly, also in these
classical field computations gluons with fairly hard momenta
$E_T/N\sim 3Q_{\rm sat}$, where $Q_{\rm sat}$ is a similar saturation
scale as $p_{\rm sat}$ above, are important for the initial production
\cite{KVn,KVet}. As this approach should coincide with the
perturbative one in the limit of high transverse momentum of produced
gluons, it is important to reduce the uncertainties related to the
possible higher-order contributions in the pQCD component. This gives
further motivation to perform the NLO computation of initial $E_T$ we
present in this paper.

In our study the NLO evaluation of the produced initial transverse
energy in central ($b=0$) $AA$ collisions is carried out by computing
the first $E_T$ moment $\sigma_{\rm QCD}\langle E_T \rangle_{\Delta Y,
p_0}$ of the perturbative $E_T$ distribution of minijets in a single
proton-proton collision within the framework of collinear
factorization.  Nuclear collision geometry is taken into account
through the standard nuclear overlap functions $T_{AA}({\bf b})$ and the
initial $E_T$ is obtained as $E_T=T_{AA}\sigma_{\rm QCD}\langle E_T
\rangle$ \cite{EKL89}.  The parameter $p_0$ defines
the extent of the perturbative treatment: we include all hard
scatterings of partons in which at least an amount of $2p_0$ of
transverse momentum is created. Note that $p_0\gg \Lambda_{\rm
QCD}$ is kept as an external parameter, allowing for different
phenomenological ways to determine its value.

We emphasize that although the initial $E_T$ in a rapidity
acceptance window $\Delta Y$ is a physical quantity, which thus can be
computed in an infrared safe manner, it is not a direct
observable. This is due to the interactions of initially produced
gluons and quarks which are expected to thermalize the QGP system
rapidly.  Due to the pressure $P$ formed, the system does work ($PdV$)
during the expansion stage, and some of the initially stopped energy
is lost. In ref. \cite{EKRT99} it was estimated that for central
collisions of $A\sim 200$ the measurable final state $E_T$ is
$E_T^{\rm initial}/3$ at the highest RHIC energy and 
$E_T^{\rm initial}/5$ at the LHC energy.

In practise, the NLO computation is performed by appealing to a
subtraction algorithm \cite{EKS89,EKS90,KS92}, applicable for a wide
class of problems requiring a NLO evaluation of a quantity fulfilling
the requirements of infrared safeness. The required
$\mathcal{O}(\alpha_s^3)$ matrix elements have been evaluated in
\cite{ES86}.  The current paper is a sequel to our previous work
\cite{ET00}, where $\sigma_{\rm QCD}\langle E_T \rangle_{\Delta Y,
p_0}$ was computed numerically and the results studied as a function
of the parameter $p_0$ and the factorization/renormalization scale
choice. The main result of \cite{ET00}, that the deviation of the NLO
results from the LO ones are not dramatically increasing as $p_0$ is
taken down from $10$ GeV to $1\dots 2$ GeV, remains also here. In
\cite{ET00} it was noted that a new and important kinematical
perturbative region has to be included in the NLO computation. Here we
study the contributions from this region in more detail.  The
relation to another recent NLO computation of the initial $E_T$,
Ref. \cite{LO98}, along with the differences from our formulation,
were also discussed in \cite{ET00}.  Now we extend the NLO study to
explore the sensitivity of the results on the choice of parton
distributions.  In addition to the dependence of the results on $p_0$
and on the scale choice, their dependence on $\Delta Y$ is explicitly
shown and the important implications discussed.  Also the effects of
nuclear parton distributions \cite{EKS98} are studied.  Finally, as an
application of the NLO computation, charged particle production for
the nucleus-nucleus collisions currently under way at RHIC as well as
for those to take place in the future at the LHC/ALICE are briefly
considered by applying the pQCD+saturation model \cite{EKRT99}.

\section{Theoretical basis of the NLO formulation} 

In the framework of independent parton-parton scatterings, the average
$E_T$ produced initially into the rapidity region $\Delta Y$ in an
$AA$ collision at an impact parameter ${\bf b}$ can be computed as
\cite{EKL89}
\begin{equation}
E_T^{AA}({\bf b},\sqrt s,p_0) = T_{AA}({\bf b})\sigma_{\rm QCD}\langle E_T
\rangle_{\Delta Y, p_0},
\end{equation}
where $T_{AA}(\bf b)$ is the standard nuclear overlap function
\cite{EKL89}.  We will focus on central collisions of large nuclei
only, for which, using the Woods-Saxon nuclear profiles,
$T_{AA}(0)\approx A^2/\pi R_A^2$.  Numerically, $T_{AA}(0)\approx
30/$mb for $A\sim200$. In what follows, we will focus on the
formulation and computation of $\sigma_{\rm QCD}\langle
E_T\rangle_{\Delta Y, p_0}$, the first moment of the minijet $E_T$
distribution in $pp$ collisions. Regarding the $E_T$ distribution in
an $AA$-collision, it is the first and also the second moments of the
$pp$-level distributions that play the key roles rather than the
distributions themselves \cite{EKL89}.

Calculation of physical cross sections in NLO pQCD is a nontrivial
task even if the scattering amplitudes are known at the desired
order. For infrared safe quantities, however, a systematic method
allowing for the cancellation of the soft and collinear singularities
in the loop and bremsstrahlung contributions exists
\cite{KS92,KUNSZT}. In this section we will briefly review this
algorithm and thoroughly describe its implementation.

\subsection{Subtraction method}

To order $\alpha_s^3$ we need to take into account processes where two
incoming hadrons produce either two or three final state partons.
Keeping the notations of \cite{KS92}, the incoming partons are
labelled by $A$ and $B$ and the outgoing partons as 1, 2 and 3.  For
the two-parton final state the appropriate kinematical variables are
$y_1, y_2, p_{T2}$ and $\phi_2$. Transverse momentum conservation
determines $p_{T1}=p_{T2}=p_T$ and $\phi_1=\phi_2+\pi$ as we do not
include any intrinsic transverse momentum.  The momentum fractions
$x_A$ and $x_B$ of the incoming partons are determined by the
conservation of energy and longitudinal momentum, resulting in
$x_{1,2}=\frac{p_T}{\sqrt s}({\rm e}^{\pm y_1}+{\rm e}^{\pm
y_2})$. Similarly, for the three-parton final state ${\bf p_{T1}} =
-({\bf p_{T2}+p_{T3}})$ and the suitable kinematical variables are
$y_1, y_2, y_3, p_{T2}, p_{T3}, \phi_2$ and $\phi_3$. The fractional
momenta become now $x_{1,2}=\frac{p_{T1}}{\sqrt s}{\rm e}^{\pm y_1}+
\frac{p_{T2}}{\sqrt s}{\rm e}^{\pm y_2} + \frac{p_{T3}}{\sqrt s}{\rm
e}^{\pm y_3}$.  An inclusive hard cross section can be written in a
general form
\begin{eqnarray}
I&=&
\int d[PS]_2 \frac{d\sigma^{2\rightarrow 2}}{d[PS]_2}S_2(p_1{^\mu},p_2^{\mu})+
\int d[PS]_3 \frac{d\sigma^{2 \rightarrow 3}}{d[PS]_3}
S_3(p_1^{\mu},p_2^{\mu},p_3^{\mu})
\nonumber \\
&\equiv & I[2 \rightarrow 2]+I[2 \rightarrow 3],
\label{Sigma1}
\end{eqnarray}
where it is implicit that the integrations take place in $4-2\epsilon$
spacetime dimensions. To condense the notation, we have denoted the
$2\rightarrow2$ and $2\rightarrow3$ differential partonic cross
sections as
\begin{equation}
\frac{d\sigma^{2\rightarrow 2}}{d[PS]_2}=\frac{d\sigma^{2\rightarrow 2}}{dp_Tdy_1dy_2d\phi_2}
\end{equation}
\begin{equation}
\frac{d\sigma^{2\rightarrow 3}}{d[PS]_3}=\frac{d\sigma^{2\rightarrow
3}}{dp_{T2}dp_{T3}dy_1dy_2dy_3d\phi_2 d\phi_3},
\end{equation}
which also defines our notation for the phase space volume elements 
$d[PS]_2$ and $d[PS]_3$.

The measurement functions $S_2(p_1^{\mu},p_2^{\mu})$ and
$S_3(p_1^{\mu},p_2^{\mu},p_3^{\mu})$ which depend on the four-momenta
of the final state partons, define the physical quantity the cross
section of which is to be computed. Such a quantity can be e.g.  a
differential one-jet cross section $\frac{d\sigma}{dp_Tdy}\big|_R$
\cite{EKS89,KS92}, or a two-jet cross section
$\frac{d\sigma}{dp_{T1}dp_{T2}dy_1dy_2}\big|_R$ \cite{2JET} for
production of observable jets with a jet cone radius $R$. In our case,
$S_2$ and $S_3$ will be designed to give
$\frac{d\sigma}{dE_T}\big|_{\Delta Y,p_0}$, the differential $E_T$
distribution of minijets which fall into a given rapidity acceptance
window $\Delta Y$ and which originate from perturbative collisions
where at least an amount $2p_0$ of transverse momentum is produced.

Let us briefly recapitulate the procedure presented in detail in
\cite{KS92} for the cancellation of singular contributions in the
above cross section.  The $2\rightarrow 2$ contribution can be written
under collinear factorization \cite{CSS85} as
\begin{eqnarray}
\nonumber \frac{d\sigma^{2\rightarrow 2}}{d[PS]_2}
&=&\frac{1}{2!}\sum_{a_A,a_B,a_1,a_2}\frac{p_{T2}}{16\pi^2s^2}
\frac{1}{x_A}\tilde f_A(a_A,x_A)\frac{1}{x_B}\tilde f_B (a_B,x_B)
\\ & & \times\langle | \mathcal{M}(a_A+a_B\rightarrow
a_1+a_2)|^2\rangle S_2(p_1^{\mu},p_2^{\mu}).
\end{eqnarray}
The incoming parton flavors are denoted by $a_A$ and $a_B$ whereas the
outgoing parton distributions are denoted by $a_1$ and $a_2$. The
matrix element squared and summed over final spin and color and
averaged over initial spins and colors is evaluated in $4-2\epsilon$
dimensions and contains the ultraviolet renormalization via the $\msbar$
scheme. The distributions $\tilde f(a,x)$ are the modified parton
distributions defined in the $\msbar$ scheme.

Similarly for the $2\rightarrow 3$ part one has
\begin{eqnarray}
\nonumber \frac{d\sigma^{2\rightarrow 3}}{d[PS]_3}
&=&\frac{1}{2!}\theta(p_{T3}<p_{T1})\theta(p_{T3}<p_{T2})
\frac{p_{T2}p_{T3}}{8(2\pi)^5s^2}\sum_{a_A,a_B,a_1,a_2,a_3}\frac{1}{x_A}f_A(a_A,x_A)\\
& &\times\frac{1}{x_B}f_B(a_B,x_B) \langle |
\mathcal{M}(a_A+a_B\rightarrow a_1+a_2+a_3)|^2\rangle
S_3(p_1^{\mu},p_2^{\mu},p_3^{\mu}),
\end{eqnarray}
where parton 3 has been identified as the one having smallest
transverse momentum, canceling a factor 3 in the original statistical
prefactor 1/3!.

The two terms, $I[2 \rightarrow 2]$ and $I[2 \rightarrow 3]$ both contain
divergent contributions proportional to $1/\epsilon$ and
$1/\epsilon^2$. All these divergent contributions cancel
remarkably against each other provided that the functions $S_2$ and
$S_3$ in the equation above, the so called measurement functions, are
infrared safe. This requirement can be stated more precisely: First,
as two of the outgoing partons become collinear, $S_3$ should reduce
to $S_2$ i.e.
\begin{eqnarray}
S_3(p_1^{\mu},(1- \lambda )p_2^{\mu}, \lambda p_2^{\mu})
&=&  
S_2(p_1^{\mu},p_2^{\mu}) ,\nonumber \\ 
S_3((1- \lambda )p_1^{\mu},p_2^{\mu}, \lambda p_1^{\mu})
&=&  
S_2(p_1^{\mu}\,p_2^{\mu}) , \nonumber \\
S_3(\lambda p_1^{\mu},(1- \lambda )p_1^{\mu},p_2^{\mu})
&=&  
S_2(p_1^{\mu},p_2^{\mu})
\label{S31} 
\end{eqnarray}
for $ 0 \leq \lambda \leq 1 $. Second, as any of the partons becomes
collinear with one of the beam momenta $ p_A^{\mu} $, $ p_B^{\mu} $ it
is required similarly, that
\begin{eqnarray}
S_3(p_1^{\mu},p_2^{\mu},\lambda p_A^{\mu})
&=&S_3(p_1^{\mu},p_2^{\mu},\lambda p_B^{\mu})=S_2(p_1^{\mu},p_2^{\mu}) ,
\nonumber \\
   S_3(p_1^{\mu},\lambda p_A^{\mu},p_2^{\mu})
&=&S_3(p_1^{\mu},\lambda p_B^{\mu},p_2^{\mu})=S_2(p_1^{\mu},p_2^{\mu}) , 
\nonumber \\
   S_3(\lambda p_A^{\mu},p_1^{\mu},p_2^{\mu})
&=&S_3(\lambda p_B^{\mu},p_1^{\mu},p_2^{\mu})=S_2(p_1^{\mu},p_2^{\mu})
\label{S32}
\end{eqnarray}

Assuming the infrared safety of the measurement functions, the
calculation based on the subtraction method \cite{KS92} then proceeds by first
separating the singular factors in $2\rightarrow 3$ matrix element
from one another. As an initial step, the matrix element is decomposed
as
\begin{equation}
\langle |\mathcal{M}|^2\rangle_{2\rightarrow 3}=\langle
|\mathcal{M}|^2\rangle_A+\langle |\mathcal{M}|^2\rangle_B+\langle
|\mathcal{M}|^2\rangle_1+\langle |\mathcal{M}|^2\rangle_2
\label{ME23}
\end{equation}
in such a way that each term $\langle |\mathcal{M}|^2\rangle_n\sim
1/(p_3\cdot p_n)$, so it contains a soft singularity for parton 3 along with
the singularity of parton 3 becoming collinear with parton $n$, where 
$n=A,B,1,2$ with $A,B$ referring to the incoming partons and 1,2 to the 
two other outgoing partons. The imposed constraint  $p_{T3}<p_{T1},p_{T2}$ 
guarantees that other singularities do not occur. Now the $2\rightarrow 3$ 
part of the cross section can be written as
\begin{equation}
I[2\rightarrow 3]=I[2\rightarrow 3]_A+I[2\rightarrow 3]_B+I[2\rightarrow 3]_1+
I[2\rightarrow 3]_2.
\end{equation}

Each of the four terms featured in the above equation can be treated
independently and in a similar manner. 
For example, the last term becomes
\begin{equation}
I[2\rightarrow 3]_2=\int \frac{d\sigma^{2\rightarrow 3}_2}{d[PS]_3}{d[PS]_3}
= \int {d[PS]_3} \frac{F_2(y_1,p_{T2},y_2,\phi_2,p_{T3},y_3,\phi_3)}
{p_{T3}[\cosh(y_2-y_3)-\cos(\phi_2-\phi_3)}, 
\end{equation}
where the divergent factor $1/(p_2\cdot p_3)$ has been expressed in
terms of the integration variables as $p_2\cdot p_3=p_{T2}p_{T3}[\cosh
(y_2-y_3)-\cos (\phi_2 -\phi_3)]$, and $F_2$ is a complicated function
containing the parton distributions, the measurement function $S_3$
and certain parts of the ($4-2\epsilon$ dimensional) squared matrix
elements due to the decomposition in Eq. (\ref{ME23}). 

The aim is now to decompose $d\sigma^{2\rightarrow 3}_2/d[PS]_3$ into
terms that are divergent but simple in a way that the integrations
over the phase space of the third parton can be performed analytically
and terms that are finite as $\epsilon \rightarrow 0$ but not so
simple. This goal is achieved by inserting zero in the following
clever manner:
\begin{eqnarray}
\nonumber
& F_2(y_1,p_{T2},y_2,\phi_2,p_{T3},y_3,\phi_3) = \\ \nonumber
& F_2(y_1,p_{T2},y_2,\phi_2,p_{T3},y_3,\phi_3) 
- F_2(y_1,p_{T2},y_2,\phi_2,0,y_3,\phi_3)\theta(p_{T3}<p_{T2}/2) \\ \nonumber
&-F_2(y_1,p_{T2},y_2,\phi_2,p_{T3},y_2,\phi_2)
+ F_2(y_1,p_{T2},y_2,\phi_2,0,y_2,\phi_2)\theta(p_{T3}<p_{T2}/2) \\ \nonumber
&+F_2(y_1,p_{T2},y_2,\phi_2,0,y_3,\phi_3)\theta(p_{T3}<p_{T2}/2) \\
&+F_2(y_1,p_{T2},y_2,\phi_2,p_{T3},y_2,\phi_2)-
F_2(y_1,p_{T2},y_2,\phi_2,0,y_2,\phi_2) \theta(p_{T3}<p_{T2}/2).
\label{subtractions}
\end{eqnarray}
The first four terms on the right-hand side combine to make up the
term $I[{\rm{finite}}]_2$ which is perfectly finite as $\epsilon
\rightarrow 0$ as both the soft singularity $p_{T3}=0$ and the collinear
singularity $3\uparrow\uparrow2$ have been subtracted.
The fifth term, containing the soft singularity, generates a term 
$I[{\rm soft}]_2$, and the remaining two terms then form the term 
$I[{\rm collinear}]_2$ which, as the name suggests, contains
the collinear singularity only. Thus 
\begin{equation}
I[2\rightarrow 3]_2= I[{\rm{finite}}]_2 + I[{\rm soft}]_2 + 
I[{\rm collinear}]_2.
\end{equation}

Due to the decomposition (\ref{ME23}) none of the
above terms contain any other singularities, but these are then
confined to other three terms $I[2\rightarrow 3]_1$, $I[2\rightarrow
3]_A$ and $I[2\rightarrow 3]_B$ defined and treated similarly.  The
procedure outlined here has therefore at this stage produced us the
decomposition
\begin{eqnarray}
\nonumber I[2\rightarrow 3]&=&\sum_{n=A,B,1,2}I[2\rightarrow 3]_n \\
&=&\sum_{n=A,B,1,2}\{I[{\rm finite}]_n+I[{\rm soft}]_n+I[{\rm
collinear}]_n\}.
\end{eqnarray}

The terms labelled `finite' above, are very complicated but finite and
can be evaluated numerically. In the singular terms $I[{\rm soft}]_2$
and $I[{\rm collinear}]_2$ the integrations over $p_{T3}$, $\phi_3$ and
$y_3$ can be performed analytically due to the simple structure of
those terms: the function $F_2$ is evaluated in each of them either on
a soft or collinear limit. After these integrations are carried out,
the resulting terms have $2\rightarrow 2$ kinematics and contain
several $1/\epsilon$ and $1/\epsilon^2$ parts. Some of these cancel
against identical (apart from the sign) terms in $I[2\rightarrow 2]$,
and the remaining parts cancel among themselves beautifully, leaving a
perfectly finite contribution behind 
- provided that the measurement functions $S_3$ reduce to 
$S_2$ as required by Eqs. (\ref{S31}) and (\ref{S32}).
The somewhat complicated structure of the
cancellation of the $1/\epsilon$ and $1/\epsilon^2$ singularities between
the $I[2\rightarrow 2]$ and $I[2\rightarrow 3]$, including the
counterterm $I^{2\rightarrow 2}[{\rm{CT}}]$ defining the $\msbar$
NLO parton distributions, is illustrated in the Table
\ref{epsilontable} below.

\begin{table}[!hbt]
\begin{center}
\begin{tabular}{|l|c|c|c|}
\hline 
 &&& \\
$\quad$Term & $1/\epsilon^2$ & $1/\epsilon$ & finite \\  
 &&& \\
\hline  &&& \\
$I_{\rmi{A,B}}^{2\rightarrow 3}[{\rm{soft}}]$  &$a_{\rmi{A,B}}$ &$b_{\rmi{A,B}}^{(1)}$; $c_{\rmi{A,B}}$ & $\times$  \\ &&& \\
$I_{\rmi{1,2}}^{2\rightarrow 3}[{\rm{soft}}]$  &$a_{\rmi{1,2}}$ &$b_{\rmi{1,2}}^{(1)};$ $c_{\rmi{1,2}}$ & $\times$  \\  &&&\\
$I_{\rmi{A,B}}^{2\rightarrow 3}[{\rm{collinear}}]$  &- &$b_{\rmi{A,B}}^{(2)};$ $c_{\rmi{A,B}};$ $d_{\rmi{A,B}}$ & $\times$  \\  &&&\\
$I_{\rmi{1,2}}^{2\rightarrow 3}[{\rm{collinear}}]$  &- &$b_{\rmi{1,2}}^{(2)};$ $c_{\rmi{1,2}}$ & $\times$  \\  &&&\\
$I^{2\rightarrow 3}[{\rm{finite}}]$ &- &- & $\times$ \\  &&&\\
$I^{2\rightarrow 2}[{\rm{Born}}]$ &- &- & $\times$ \\  &&&\\
$I^{2\rightarrow 2}[{\rm{HO}}]$ &$a_{\rmi{1,2}};$ $a_{\rmi{A,B}}$ &$b_{\rmi{A,B}}^{(1)};b_{\rmi{A,B}}^{(2)};$ $b_{\rmi{1,2}}^{(1)};b_{\rmi{1,2}}^{(2)}$ & $\times$ \\  &&&\\
$I^{2\rightarrow 2}[{\rm{CT}}]$ &- &$d_{\rmi{A,B}}$ &- \\  &&&\\
\hline
\end{tabular}
\end{center}
\caption{\small The singularity content of the different terms contributing
to the cross section (\ref{Sigma1}). The top panel specifies the type
of the singularity and the symbols $a_i, b^{(1)}_i, b^{(2)}_i,c_i,d_i$
($i=A,B,1,2$) represent the coefficient functions of the singularities
of that type. Singularities between the different terms cancel
whenever two identical symbols appear. For instance, the
$1/\epsilon^2$ singularity in the term $I_{\rmi{A,B}}^{2\rightarrow
3}[{\rm{soft}}]$ is cancelled by that contained in the NLO
$2\rightarrow2$ term $I^{2\rightarrow 2}[{\rm{HO}}]$. The counterterm
$I^{2\rightarrow 2}[{\rm{CT}}]$ defines the NLO parton distributions
in the $\msbar$ scheme. Appearance of a finite contribution to the
cross section (\ref{Sigma1}) in each term is also indicated by the 
crosses in the last column.}
\label{epsilontable}
\end{table}

The simple result of the whole procedure then is that the full cross
section can be written as \be I=I[2 \rightarrow 2\, {\rm net}]+I[2
\rightarrow 3\, {\rm net}] \ee where the `net' contributions are
obtained by removing all of the $1/\epsilon$ and $1/\epsilon^2$ terms
and letting $\epsilon \rightarrow 0$. Due to the infrared safety of
this NLO algorithm the divergences originating from the higher order
loop diagrams to $2\rightarrow 2$ scatterings and soft particle
emission of $2\rightarrow 3$ scatterings cancel each other exactly
\cite{KS92,KUNSZT}.

\subsection{Measurement function for the minijet $E_T$}

We now turn to the detailed description of the implementation of the
algorithm discussed in the previous section. The main points have been
addressed already in our previous work \cite{ET00} but for
completeness, we wish to do this again in here. The quantity we are
first after, is the total $E_T$ carried by minijets into a chosen
rapidity acceptance window in an average inelastic $pp$ collision.
Furthermore, in order for the perturbative treatment to be valid, the
partonic collisions must be required to be sufficiently hard. All this
has to be included in the measurement function which must be infrared
safe by construction.

For the computation of observable jets, the starting point would be to
define the jet cone, i.e. the definition of when two nearly collinear 
final state partons are to be considered as  one jet and when
as as two separate jets. Such a measurement function is not applicable
here but an acceptance region of a new type must be defined.

\begin{figure}[htb]
\vspace{-2cm}
\begin{center}
\includegraphics[width=12cm,trim=0 280 0 0]{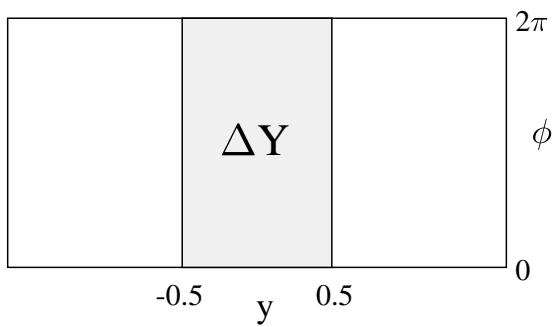}
\end{center}
\caption{\small The acceptance region $\Delta Y$ in the plane of
rapidity $y$ and azimuthal angle $\phi$. The window $\Delta Y$ is centered
around $y=0$ here.
}
\label{ycut}
\end{figure}

In a hard scattering of partons in the NLO, we may have one, two,
three or zero minijets in our rapidity acceptance region $\Delta Y$,
defined in the $(y,\phi)$-plane as
\begin{equation}
\Delta Y=\{(y,\phi) : y_{min}\le y \le y_{max},\quad 0 \le \phi \le 2 \pi\}.
\end{equation}
In our case, $\Delta Y $ will always be centered at $y=0$, see
Fig. \ref{ycut}.  As only massless partons are considered here, the
transverse energy entering $\Delta Y$ can be defined as a sum of the
absolute values $p_{Ti}$ of the transverse momenta of those partons
whose rapidities are within $\Delta Y$:

\begin{equation}
E_T = \epsilon(y_1)p_{T1} + \epsilon(y_2)p_{T2} + \epsilon(y_3)p_{T3},
\label{ET}
\end{equation}
where the step function $\epsilon(y_i)$ is defined as in \cite{EKL89},
\begin{equation}
\epsilon(y_i) \equiv \left\{
\begin{array}{ll}
        1 & \mbox{if $y_i\in\Delta Y$}\\
        0 & \mbox{otherwise}.
\end{array}
\right.
\end{equation}

Since we cannot solve full QCD, but have to resort to perturbative
treatment, we must carefully specify which collisions are to be considered 
perturbative, i.e. hard enough. We define the perturbative collisions to be 
those with a large enough amount ($\gg \Lambda_{\rm QCD}$) of transverse 
momentum produced, regardless of where the partons go in rapidity. For the 
different final state kinematics, this implies
\begin{eqnarray}
2\rightarrow2:& p_{T1}+p_{T2}\ge  2p_0 \nonumber \\
2\rightarrow3:& \quad\quad\quad\quad p_{T1} + p_{T2} + p_{T3}\ge 2p_0, \quad p_0\gg
\Lambda_{\rm{QCD}} \label{pert}
\end{eqnarray}
where the parameter $p_0$, which restricts the
computation below, is a fixed external parameter which does not depend
on $\Delta Y$.  For $2\rightarrow2$ processes, appearing both in LO
and in NLO, where the scattered partons are back-to-back in the
transverse plane, the condition above means that individual minijets,
inside or outside $\Delta Y$ always have $p_T\ge p_0$. For the
$2\rightarrow3$ processes, which appear only in NLO, a single parton
may have also less than $p_0$ transverse momentum as long as the total
transverse momentum released in the partonic process exceeds $2p_0$.
This feature, which is also the main difference to the formulation of the same
problem in Ref. \cite{LO98}, will have interesting implications later on.
We will investigate the dependence of the results on $p_0$ and whether
the NLO results remain stable relative to LO as $p_0$ is taken into
the few-GeV realm. To estimate the transverse energy produced in $AA$
collisions at RHIC and LHC, additional phenomenology (see e.g
\cite{EKRT99}) needs to be introduced. This phenomenology will be
separately discussed in Sec. 4.

The infrared safe measurement functions $S_2$ and $S_3$ \cite{KS92} can now
be written down by combining the definitions of perturbativeness of
the collisions and the definition of $E_T$ in $\Delta Y$. For $2\rightarrow2$
scatterings, we define
\begin{equation}
S_2(p_1^{\mu},p_2^{\mu}) =  
\Theta(p_{T1}+p_{T2}\ge 2p_0)
\delta(E_T-[\epsilon(y_1)p_{T1}+\epsilon(y_2)p_{T2}])
\label{S2}
\end{equation}
and  for $2\rightarrow 3$ scatterings correspondingly
\begin{equation}
S_3(p_1^{\mu},p_2^{\mu},p_3^{\mu}) = 
\Theta(p_{T1}+p_{T2}+p_{T3}\ge 2p_0)
\delta(E_T-[\epsilon(y_1)p_{T1}+\epsilon(y_2)p_{T2}+\epsilon(y_3)p_{T3}]).
\label{S3}
\end{equation}
where $\Theta$ is a step function.

It is easy to verify that these measurement functions indeed fulfill
the requirements (\ref{S31}) and (\ref{S32}) for infrared safety. This
is also intuitively clear, as the total $E_T$ is not sensitive to
whether it is carried by one parton or two collinear partons, or by two
partons of which one is at the soft limit.

The measurement functions above define the semi-inclusive $E_T$ distribution 
of minijets in $\Delta Y$ in $pp$ collisions, introduced in LO in \cite{EKL89}.
Extending this now to NLO gives
\begin{eqnarray}
\frac{d\sigma}{dE_T}\bigg|_{p_0,\Delta Y} 
&=& 
\frac{d\sigma}{dE_T}\bigg|_{p_0,\Delta Y}^{2\rightarrow2} +
\frac{d\sigma}{dE_T}\bigg|_{p_0,\Delta Y}^{2\rightarrow3} 
\nonumber \\
&=& 
\int d[PS]_2 \frac{d\sigma^{2\rightarrow 2}}{d[PS]_2}S_2(p_1^{\mu},p_2^{\mu}) +
\int d[PS]_3 \frac{d\sigma^{2 \rightarrow 3}}{d[PS]_3}S_3(p_1^{\mu},p_2^{\mu},p_3^{\mu}),
\nonumber \\
\label{dET}
\end{eqnarray}
with the notation as in Eq. (\ref{Sigma1}).

The minijet $E_T$ distribution above is normalized to the integrated
cross section $\sigma_{\rm jet}(\sqrt s,p_0)$, as explained in
\cite{EKL89}. As a side remark, we note that although we do not
explicitly consider the computation of $\sigma_{\rm jet}$ here, it is
an infrared safe quantity to compute, i.e. the extension of its definition 
to NLO can be done. The main interest now, as
already emphasized, is to study the behaviour of the first moment of
the $E_T$ distribution (\ref{dET}) as a function of the parameter
$p_0$ which defines the extent of our perturbative calculation.
Integrating the delta functions away in Eq.~(\ref{dET}) we get
\begin{eqnarray}
\sigma_{\rm QCD}\langle E_T\rangle_{\Delta Y,p_0} \equiv
\int_0^{\sqrt s} dE_T\,E_T \frac{d\sigma}{dE_T}\bigg|_{p_0,\Delta Y} 
= \sigma \langle E_T\rangle_{\Delta Y,p_0}^{2\rightarrow2}
+ \sigma \langle E_T\rangle_{\Delta Y,p_0}^{2\rightarrow3},
\label{sET}
\end{eqnarray}
where 
\begin{eqnarray}
\sigma \langle E_T\rangle_{\Delta Y,p_0}^{2\rightarrow2} &=&
\int d[PS]_2 \frac{d\sigma^{2\rightarrow 2}}{d[PS]_2}
\tilde S_2(p_1^{\mu},p_2^{\mu}) \\
\sigma \langle E_T\rangle_{\Delta Y,p_0}^{2\rightarrow3} &=&
\int d[PS]_3 \frac{d\sigma^{2 \rightarrow 3}}{d[PS]_3}
\tilde S_3(p_1^{\mu},p_2^{\mu},p_3^{\mu}).
\label{sET3}
\end{eqnarray}
The measurement functions for the first $E_T$-moment above are denoted by   
\begin{eqnarray}
\tilde S_2(p_1^{\mu},p_2^{\mu}) &=& \bigg[\epsilon(y_1)+\epsilon(y_2)\bigg]p_{T2}
\Theta(p_{T2}\ge p_0) \\ 
\tilde S_3(p_1^{\mu},p_2^{\mu},p_3^{\mu}) &=& 
\bigg[\epsilon(y_1)p_{T1}+\epsilon(y_2)p_{T2}+\epsilon(y_3)p_{T3}\bigg]
\Theta(p_{T1}+p_{T2}+p_{T3}\ge 2p_0),
\label{S3tilde}
\end{eqnarray}
where $p_{T1}=p_{T2}$ in the former term and $p_{T1}= |{\bf
p}_{T2}+{\bf p}_{T3}|$ in the latter one.  Naturally, also $\tilde
S_2$ and $\tilde S_3$ fulfill the criteria (\ref{S31}) and
(\ref{S32}), which ensures that $\sigma \langle E_T\rangle_{\Delta
Y,p_0}$ is a well-defined infrared safe quantity to compute. From the
general form of the equations above, we notice that, by replacing the
measurement functions $S_2$ and $S_3$ in Eq. \ref{Sigma1} by $\tilde
S_2$ and $\tilde S_3$ defined above, we can exactly follow the
formulation and solution of the problem as given in ref. \cite{KS92}
and summarized in the previous section.

Finally, apart from the measurement functions themselves but related to the 
problem being infrared safely defined, the renormalization scale $\mu_R$ in 
the strong coupling $\alpha_s(\mu_R)$ and the factorization scale $\mu_F$
in the parton distributions $f_i(x,\mu_F)$ have to be chosen in such a way 
that the scales for the $2\rightarrow3$ terms reduce to those for the terms 
with $2\rightarrow2$ kinematics in the soft and collinear limits.

\section{Numerical Evaluation}

The formulation of the minijet $E_T$ production thus exists. To
study its implications and to estimate its usefulness a numerical
study has to be carried out next.

Due to the simple structure of the two-particle phase space, one can 
analytically describe the kinematically allowed region with the cuts
imposed by $\Delta Y$. The part consisting of two-particle final states 
is therefore easily evaluated with a high precision by using the available 
integration routines of the fortran NAG-library \cite{NAG}.

The seven-dimensional phase space integral over the three-particle
phase space reduces to a six-dimensional one after the symmetry of the
measurement function $S_3$ under the azimuthal angle $\phi_2$ is taken
into account.  In principle, these six-dimensional integrals could be
treated similarly as the three-dimensional ones over the two-particle
phase space, but due to the additional kinematical cuts introduced by
the subtraction terms (the additional step functions in
Eq. (\ref{subtractions})) this does not work out in practice.  Rather
one is forced to use fairly crude and quite time-consuming Monte Carlo
integration. We have used another NAG subroutine to take care of these
terms.  Fortunately, the major part of the cross sections comes from
the Born level terms plus the finite parts of the subtraction terms
having $2\rightarrow 2$ kinematics. The finite terms with
$2\rightarrow 3$ kinematics are smaller and therefore their
contribution to the total relative numerical error of the results is
estimated to be rather small, below four per cent in any case.

In the numerical evaluation of the integrands with $2\rightarrow3$ we
are able to use certain subroutines of the jet-program of Ellis,
Kunszt and Soper \cite{PROGRAM}. Also the partonic book-keeping is
taken care of by adopting the method and  the permutation tables of different
subprocesses used in the program of EKS. The number of flavors is
fixed to be $N_f = 4$, and calculations were performed using two
different sets of parton distribution functions (PDF): GRV94 \cite{GRV94}
and a more recent set CTEQ5 \cite{CTQ5}. For $\Lambda_{\rm QCD}$ we
take the value as quoted in the PDF set used.

Before the numerical evaluation, also the renormalization and
factorization scales have to be fixed. In principle, the dependence of
the results on $\mu_R$ and $\mu_F$ should be studied independently. In
practise, however, the Monte Carlo integrals being tedious to solve
numerically, we follow the common practise and choose these to be
equal, $\mu_R=\mu_F\equiv \mu$, and study the dependence of the
results on $\mu$ only.  Regarding the scale itself, the most physical
choice should reflect the perturbativeness of the collision, therefore
we set $\mu$ proportional to the total transverse momentum produced in the
hard process, regardless whether the partons are in $\Delta Y$ or not,
\begin{eqnarray}
2\rightarrow2: \quad \mu&=&N_{\mu}\times \frac{1}{2}(p_{T1}+p_{T2})= N_{\mu}\times p_T \\ \nonumber
2\rightarrow3: \quad \mu&=&N_{\mu}\times \frac{1}{2}(p_{T1}+p_{T2}+p_{T3}),
\label{scalechoice}
\end{eqnarray}
where $N_{\mu}$ is a constant of the order of unity.
This choice is infrared safe in the sense of Eqs. (\ref{S31})-(\ref{S32}),
as is required for the exact cancellation of the divergent terms.
We also note that with this scale choice one is able to include the new 
kinematical region, where two partons fall outside $\Delta Y$ and 
one with small momentum ($p_T<p_0$) falls inside, contributing thus to
the region $E_T<p_0$ not present in the computation of the perturbative 
$2\rightarrow 2$ processes.

As there is some freedom in choosing the scale, other possibilities
can also be thought of. For example in Ref. \cite{LO98} the total
$E_T$ entering $\Delta Y$ was considered. This, we think, leads to an
ambiguity already at the $2\rightarrow2$ level, as almost identical
hard subprocesses are given very different scales depending whether
both of the partons fly into $\Delta Y$ or not. This issue is also clearly
connected with the definition of the perturbativity of the partonic
processes, i.e. which collisions are to be included in the computation
and which excluded. For high-$p_T$ jets, of course, $\mu\sim
p_T^{\rm{jet}}$ is the most natural scale choice, but in our case we
would like to emphasize the semi-inclusive nature of the problem, and
also that the minijet $E_T$ distribution is, unfortunately, not a
direct observable, especially not in $AA$ collisions. 

\section{Results}

\subsection{Dependence on $p_0$ and on the PDF set}

The main goal, for the reason of the importance of the semihard region
in initial $E_T$ production in $AA$ collisions, is now to study the
behaviour of $\sigma\langle E_T\rangle$ as the parameter $p_0$
controlling the perturbativity is taken from large scales into the
few-GeV domain. On the basis of high-$p_T$ inclusive jet computations
\cite{EKS89,EKS90,EW_HPC} one may well expect fairly large NLO
contributions: at scales of few tens of GeV, say, NLO improved results
may well be larger by a factor of two as compared to LO ones. This
largeness in the absolute magnitude of the correction terms does not
imply that the perturbation series would be in a stall well out of the
region of convergence, but rather is a signal of the possibility of
some new element becoming manifest only at such high order. For the
case of observable jets, the new element is the dependence of the
NLO cross sections on the jet cone $R$. Another well-known example of
precisely such a case is the Drell-Yan dilepton production where the
NLO corrections bring in QCD as a new element \cite{DY}.  In this
sense, the question of the convergence of the perturbation series is
moved to the behaviour of the next-to-next-to-leading order terms.

In our case the new element the NLO definition of the problem
introduces, is the appearance of a new kinematical region: for
$2\rightarrow2$ processes, the $E_T$ distribution is empty at scales
$E_T<p_0$ but not anymore so in the $2\rightarrow3$ case. The crucial
question therefore concerns the behaviour of the results as $p_0$ is
decreased from, say, 10 GeV down to to 1...2 GeV. If the NLO result is
of a similar magnitude as at 10 GeV, as will be shown to be the case,
the conclusion is that the NLO result is stable and pQCD can be
appealed to at such small scales. Another interesting point, already
studied in \cite{ET00}, is the dependence of $\sigma\langle
E_T\rangle$ on the choice for the renormalization and factorization
scale, and whether this behaviour substantially deviates from that in
LO.

To study the stability in the sense described above we define two
$K$-factors: 
\begin{equation}
K'=\frac{\rm{NLO}}{\rm{LO'}}, \quad\quad K=\frac{\rm{NLO}}{\rm {LO}},
\label{Ks}
\end{equation}
where LO$'$ is the first term of the NLO expansion, i.e. it is the
Born term evaluated with 2-loop $\alpha_s$ and NLO parton distribution
functions. Unprimed LO stands for leading order result evaluated with
1-loop $\alpha_s$ and LO parton distribution functions, this being the
``truly'' leading order result. These two $K$-factors tell different
tales: The unprimed $K$ is the one commonly introduced in the LO
computations to bring the elements of NLO in. The $K'$ on the other
hand measures roughly the difference between two subsequent terms in
the perturbation series, and possibly carries somewhat more
information about the actual convergence of the perturbation series,
since in $K'$ the same parton distributions appear both in the
nominator and denominator. Which one is more relevant, is somewhat of
a matter of taste, though, and this is why both $K$-factors are
discussed.

\begin{figure}[hbt]
\vspace{-0cm}
\begin{minipage}[b]{0.5\linewidth}
	\centering \includegraphics[width=8cm,trim=100 100 0
	0]{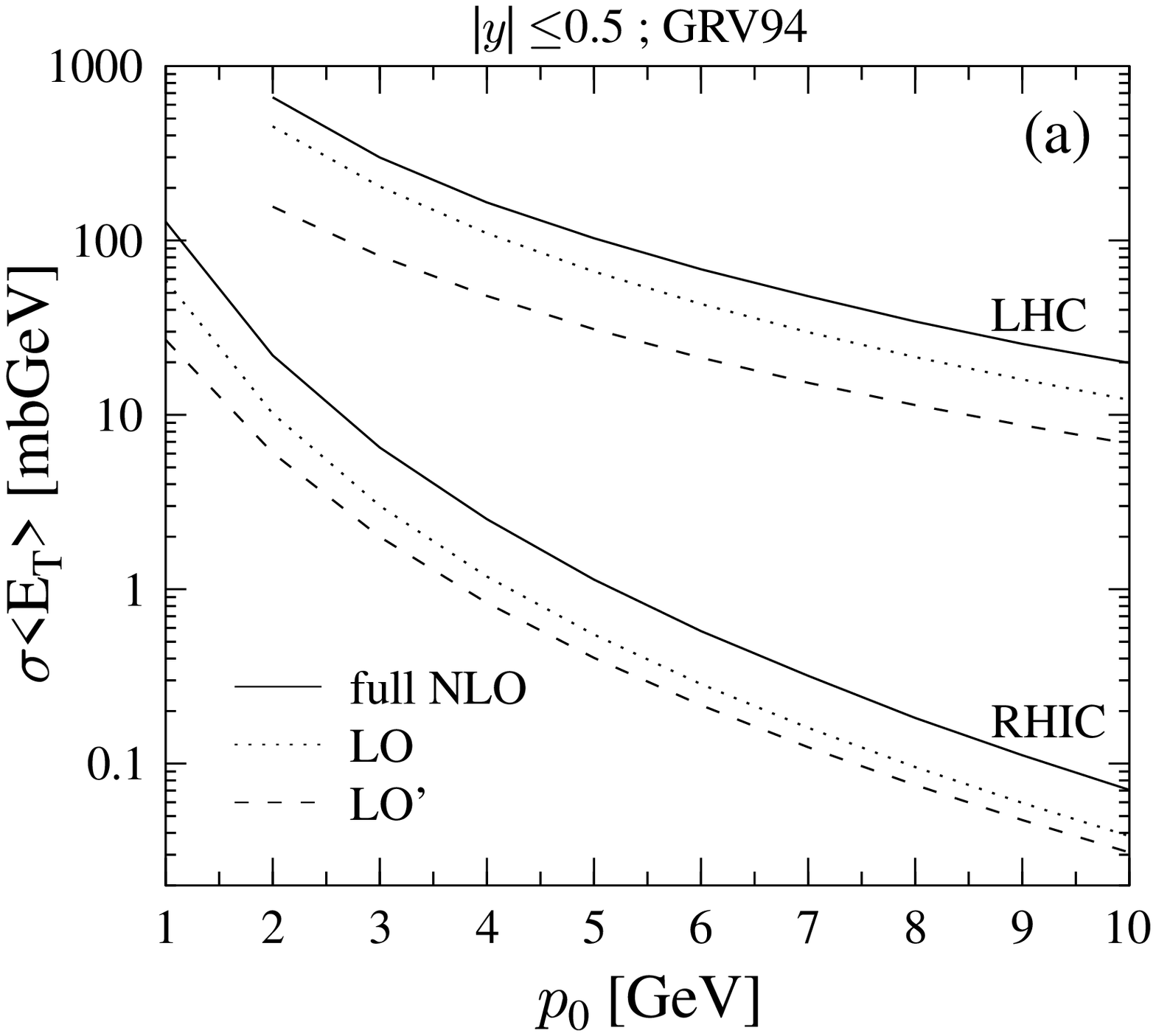} 
\end{minipage}
\hspace{1cm}
\begin{minipage}[b]{0.5\linewidth}
	\centering \includegraphics[width=8cm,trim=100 100 0
	0]{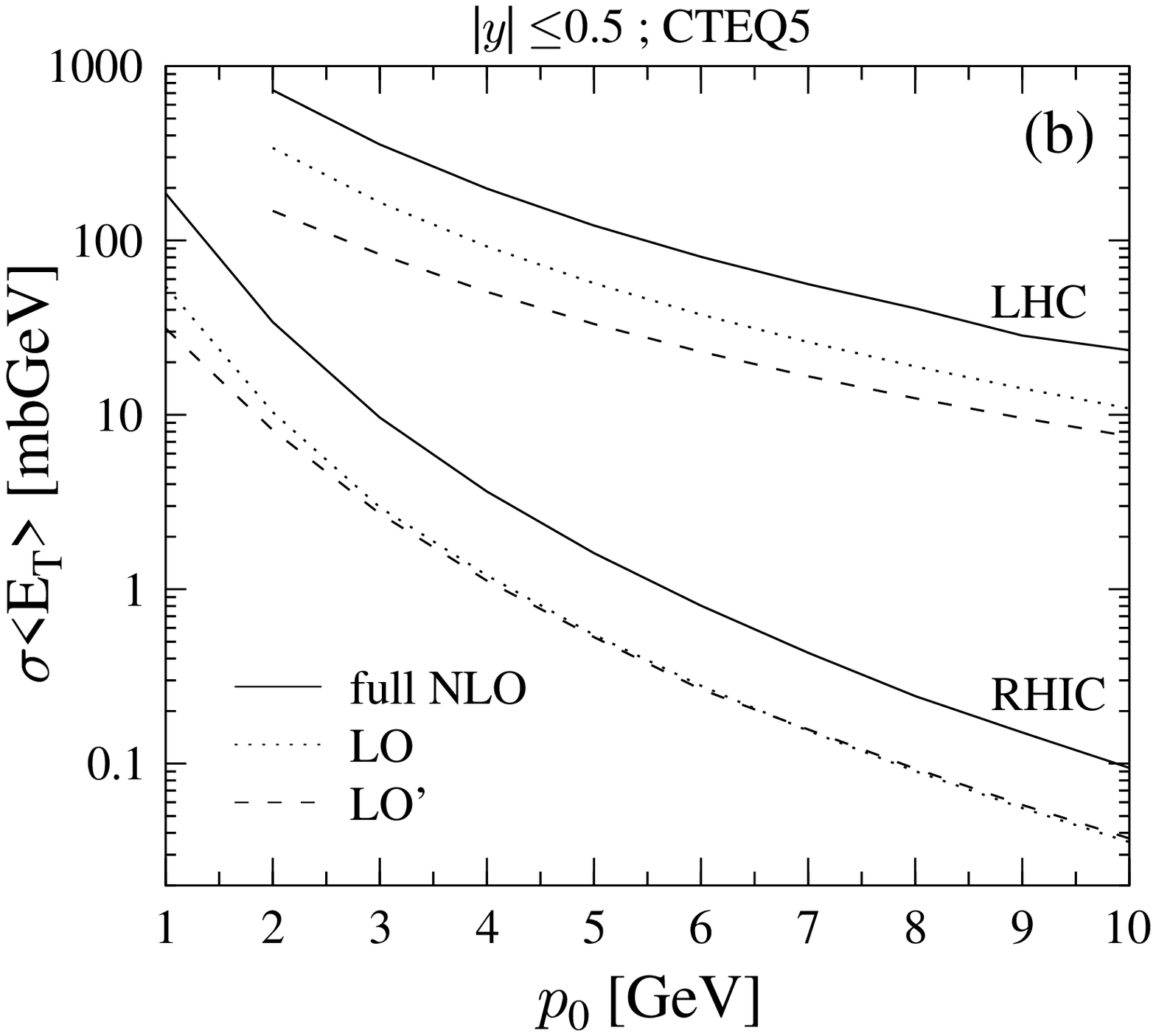} 
\end{minipage}
\caption{ \small (a) The first moment $\sigma\langle E_T \rangle$ of the 
minijet $E_T$ distribution as a function of the minimum transverse 
momentum scale $p_0$.  Three curves for both RHIC, $\sqrt s=200$ GeV, 
and LHC, $\sqrt s=5500$ GeV, are shown: the solid lines are the full 
NLO results computed with the GRV94-HO parton distributions,
the dashed lines are the LO results evaluated with 2-loop $\alpha_s$ 
and NLO parton distributions, and the dotted lines in between are the
truly LO results evaluated with 1-loop $\alpha_s$ and GRV94-LO parton
distributions. The scale choice for all points in this figure
is $N_{\mu}=1$ in Eq. (\ref{scalechoice}).
(b) The same but computed with the CTEQ5M and CTEQ5L parton distributions.
Labelling of the curves is same as in panel a.}  
\label{p0dep}
\end{figure}
\vspace{1cm}

In Fig. \ref{p0dep}a the results for $\sigma\langle E_T \rangle$
computed with GRV94 parton distribution functions are shown as a
function of the parameter $p_0$. The solid curves are the NLO results,
the dotted and dashed ones show the LO results computed in the two
different ways described above. In Fig. \ref{p0dep}b we repeat the
calculation by using the CTEQ5 parton distribution functions. The
corresponding $K$-factors ($K$ and $K'$) are collected in Table
\ref{Ktable} for some values of $p_0$ from both figures.  The
immediate observation from the figures and the table is that the full
NLO results deviate very systematically from the LO ones: Both $K$ and
$K'$ increase as $p_0\rightarrow 1$ GeV, but the factor $K$ appears to
increase less than the factor $K'$. The steady increase in $K'$ can be
understood as a signal of the nearby borderline of pQCD. In a more
practical sense the differences between LO and LO$'$ arise at RHIC
energies from the differences in $\alpha_s$ and $\Lambda_{\rm{QCD}}$
and at LHC energies, in addition to the previous, from the difference
between the LO and NLO parton distributions.

Based on the more stable behaviour of $K$, it may well be argued that
if one trusts the perturbative computation at 10 GeV scale, one can
equally well count on it also at 2 or 3 GeV scales.
Another observation is that the $K$-factors do depend on the PDF sets
chosen. 


\begin{table}[!hbt]
\begin{center}
\begin{tabular}{|l|c|c||c|c|}
\hline $p_0$ & $K_{RHIC}^{\prime GRV(CTEQ5)}$ & $K_{RHIC}^{GRV(CTEQ5)}$ &
$K_{LHC}^{\prime GRV(CTEQ5)}$ & $K_{LHC}^{GRV(CTEQ5)}$ \\ 
\hline 
1  & 4.8 (5.9) & 2.2 (3.4) & -         & -         \\
2  & 3.6 (4.2) & 2.2 (3.3) & 4.2 (4.9) & 1.5 (2.1) \\ 
4  & 3.0 (3.2) & 2.1 (3.0) & 3.4 (3.9) & 1.5 (2.2) \\
6  & 2.7 (3.0) & 2.0 (2.9) & 3.2 (3.5) & 1.6 (2.2) \\
8  & 2.4 (2.6) & 1.9 (2.7) & 3.0 (3.3) & 1.6 (2.2) \\
10 & 2.3 (2.5) & 1.8 (2.7) & 2.9 (3.1) & 1.6 (2.2) \\
\hline
\end{tabular}
\end{center}
\caption{\small Values of the two $K$-factors defined in the text for different
values of $p_0$. The numbers are shown for both RHIC and LHC energies and
for two different PDF-sets, the GRV94 and CTEQ5.}
\label{Ktable}
\end{table}

It should be emphasized once more, that here $p_0$ is to be strictly
regarded as an external parameter, on which the dependence of the
results is studied. To form the initial $E_T$ in $AA$ at RHIC and at
the LHC, one has to decide upon at which value of $p_0$ the
$\sigma\langle E_T\rangle$ should be taken. The expectation is that
$p_0\sim 1\dots 2$ GeV, possibly depending on $\sqrt s$ and $A$
\cite{EKRT99}, but precise estimation of this value is beyond the
perturbative approach in the sense that it requires introduction of
further phenomenology or knowledge of non-perturbative physics
involved.  Regardless of the actual mechanism to determine the
suitable value of $p_0$ the basic question of applicability of pQCD in
the few-GeV domain remains, and it is precisely this question that is
currently being answered.

\subsection{Dependence on the scale choice}

\begin{figure}[hbt]
\vspace{-1cm}
\begin{center}
\hspace{-1cm}
\begin{minipage}[b]{0.5\linewidth}
	\centering \includegraphics[width=10cm,trim=50 100 0
	0]{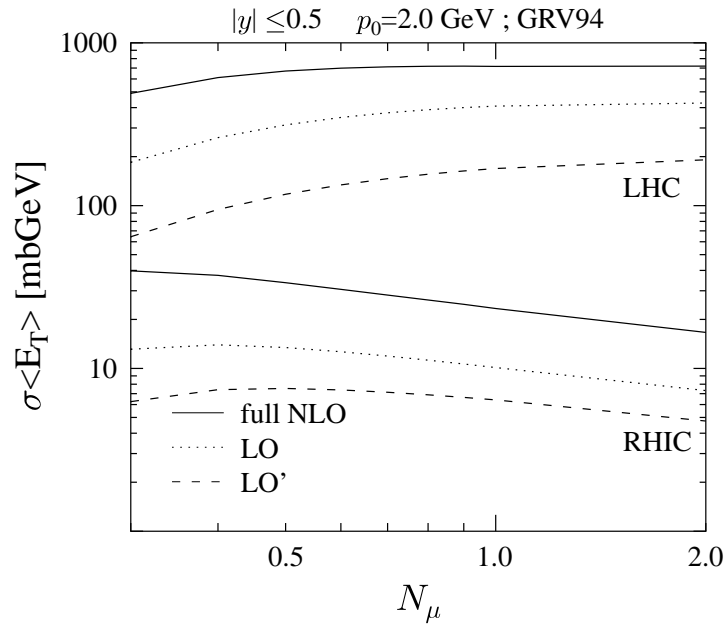} 
\end{minipage}
\end{center}
\caption{\small
The first moment $\sigma\langle E_T \rangle$ as a function of 
$N_{\mu} = \mu/\sum_i(p_{Ti})$ (see Eq. (\ref{scalechoice}))
at a fixed value of $p_0=2$ GeV. The labelling of the curves is 
same as in Figs. \ref{p0dep}, and the GRV94 parton distributions
have been used.}  
\label{GRVmudep}
\end{figure}

The dependence of $\sigma\langle E_T \rangle$ on the
renormalization/factorization scale $\mu$, chosen according to in
Eq. (\ref{scalechoice}), has essentially been presented in Ref.
\cite{ET00} but for completeness let us discuss this a bit further
here.  In Fig. \ref{GRVmudep}, we plot $\sigma\langle E_T \rangle$ at
a fixed value $p_0=2$ GeV as a function of the $N_{\mu}$ in
Eq. (\ref{scalechoice}).  The results in Fig. \ref{p0dep}(a) for
this $p_0$ are found at $N_{\mu}=1$ in the figure.

At RHIC the results depend somewhat on the chosen scale. This
behaviour is as what could be expected on the basis of the studies of
cross sections of observable jets \cite{EKS90,EW_HPC}, where it is
more difficult to find a plateau as a function of $\mu/p_T$ towards
smaller jet momenta.  At the LHC $\sigma\langle E_T \rangle$ exhibits
a clear plateau around $N_{\mu}\sim 1$. Similar behaviour is observed
already in both leading order curves, due to cancellation between the
rise of gluon distributions and the decrease of $\alpha_s$ with an
increasing scale.  Therefore the plateau cannot be attributed to a
fast convergence of the perturbation series but the NLO results depend
equally strongly (RHIC) or weakly (LHC) on the scale choice as the LO
results do.  Another consequence of this is that the $K$-factor is
close to constant near $N_{\mu}\sim 1$.


\subsection{A new kinematical region}

As noted already in our earlier work \cite{ET00}, the NLO computation
allows for a new kinematical region to be included.  The existence of
such a new contribution arises in the following way: in LO and in the
loop-corrected NLO terms, where only $2 \rightarrow 2$ processes are
considered, the two final state partons are produced back-to-back with
equal transverse momenta. The smallest amount of $E_T$ entering the
acceptance region $\Delta Y$ is therefore equal to $p_0$, and
consequently the minijet $E_T$-distribution is completely empty below
$E_T=p_0$. In NLO, however, the $2\rightarrow 3$ scatterings are
included and the situation where one of the three partons has
transverse momentum smaller than $p_0$ but the other two have larger
momenta, as required by the condition $p_{T1}+p_{T2}+p_{T3}\geq 2p_0$
and momentum conservation, may arise. Now, the two partons having
transverse momenta larger than $p_0$ may fall outside the acceptance
window and only the third one with $p_T$ equal to a fraction of $p_0$
enters $\Delta Y$, thus adding a contribution from the domain $E_T\leq
p_0$ into the $E_T$-distribution.

\begin{figure}[hbt]
\vspace{-1.0cm}
\begin{center}
\hspace{-1cm}
\begin{minipage}[b]{0.5\linewidth}
	\centering \includegraphics[width=10cm,trim= 50 100 0
	0]{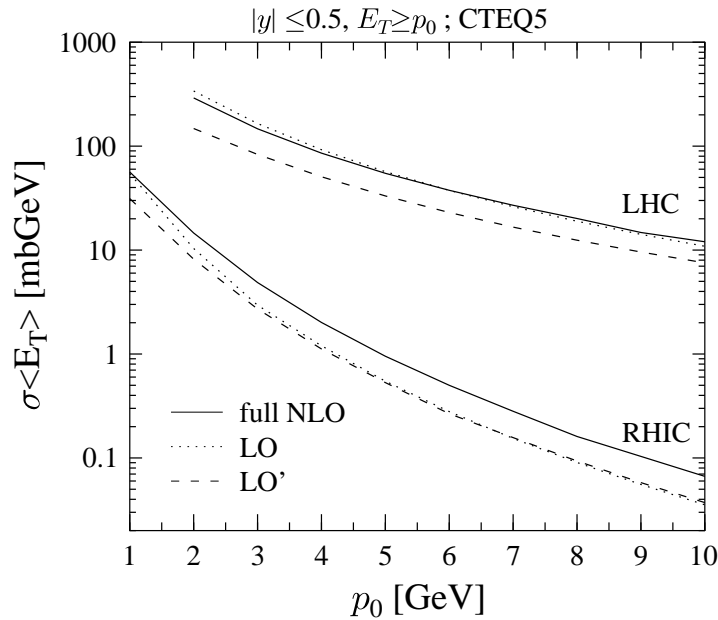} 
\end{minipage}
\end{center}
\caption{(a) \small The quantity $\sigma\langle E_T \rangle$ as a
function of $p_0$, computed as in Fig. \ref{p0dep}(b) but with an
additional cut $\theta(E_T\ge E_0=p_0)$). Labelling of the curves is
the same as in Figs. \ref{p0dep}, and CTEQ5 partons distributions are used. 
Note that the LO curves (dotted and dashed) are identical with those 
in Fig. \ref{p0dep}(b).}
\label{newregion}
\end{figure}

To identify the contribution from the new domain at $E_T<p_0$ from the
results shown in Figs. \ref{p0dep}, we compute $\sigma\langle E_T
\rangle$ by performing the integration in Eq.  (\ref{sET}) from
$E_T=p_0$ to $\sqrt s$ instead of integrating all the way from 0. To
the measurement function (\ref{S3tilde}) this adds a term
$\Theta(E_T\ge p_0)$. This is actually close to the approach of
Ref. \cite{LO98} (but still not the same, as our $p_0$ is a constant
independent of $\Delta Y$), where the new kinematic region is not
included.  The obtained results are shown in Fig \ref{newregion}, and
the value of the resulting $K'$-factor is comparable with the ones
obtained in \cite{LO98}.


Comparing this figure with Fig. \ref{p0dep}b, one observes that the
$K$-factors come down by a factor of two, hence enhancing the
indications of the validity of the perturbative treatment. At the LHC
energy, and at RHIC for $p_0\sim 1$ GeV, the NLO result shows
practically no deviation from the truly LO one. Also, by comparison of
the two figures, the slight growth of the $K$ factors, especially
relative to LO$'$, can now be attributed to the inclusion of the new
kinematical region at $E_T<p_0$. The conclusion thus is that a new
kinematical domain at $E_T\le p_0$ not only exists, but it is also an
important new feature in the extension of the problem to NLO. This
also hints that if a convergence of the perturbation series is looked
for in the semihard region, at least a NNLO computation should be studied.

It is also interesting to note that inclusion of the new kinematical domain
brings the perturbative approach towards the studies of
the classical field configurations
\cite{KMcLW,GB,GMcL,GUO}, where the dominant contribution is
obtained from the process of two valence quarks exchanging a gluon and
emitting a gluon. All such three-particle processes, and most importantly 
the purely gluonic ones,  are now rigorously included in our computation, 
supplemented of course with the criterion for perturbativity, i.e. that
in our approach the partons which scatter forward and backward outside 
$\Delta Y$, are required to carry large enough $p_T$.

\subsection{Dependence on $\Delta Y$}

\begin{figure}[hbt]
\vspace{-1.0cm}
\begin{center}
\hspace{-1cm}
\includegraphics[width=10cm,trim=50 100 0
	0]{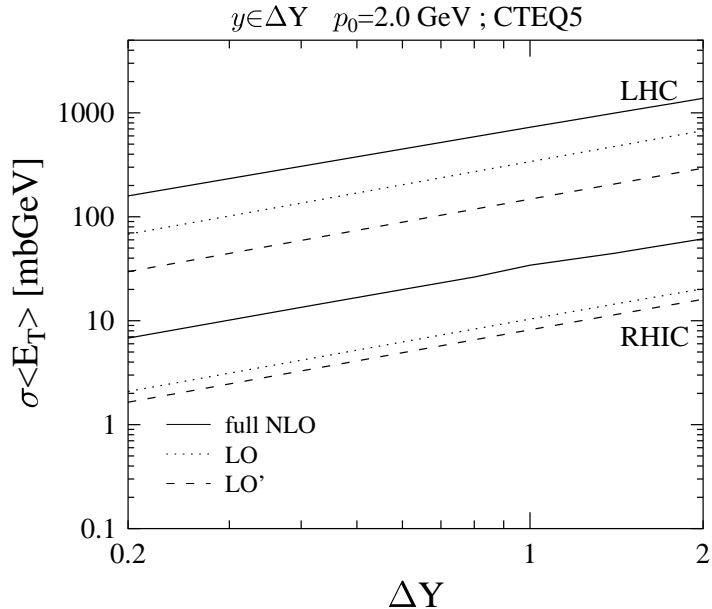}
\caption{\small The quantity $\sigma\langle E_T \rangle$ as a function of
the rapidity acceptance window $\Delta Y$ with a fixed value of $p_0=2$ GeV.
The three lower curves are for RHIC and the three topmost curves for the LHC. 
As before, the solid curves stand for the NLO results, and the dotted and dashed ones for the LO results.
}
\label{dydepCTQ}
\end{center}
\end{figure}

So far all the results presented have been evaluated for the central
rapidity unit, and it would certainly be desirable to know how the
results change as the rapidity interval $\Delta Y$ (centered around
$y=0$) is varied. This is shown in  Fig. \ref{dydepCTQ}, where
$\sigma\langle E_T \rangle$ is plotted as a function of $\Delta Y$, 
keeping $p_0$ at a fixed value of 2 GeV. The results are seen to be 
nearly linear in $\Delta Y$ around $\Delta Y=1$.

In fact this is an important result as it increases confidence in
computing the initial conditions from pQCD \cite{EKRT99}
in the following way:  The average initial energy densities in the
central spatial region, $z\sim 0$, can be estimated through the
Bjorken formula \cite{BJORKEN}
\begin{equation}
\epsilon_i = \frac{E_T^{AA}({\bf{b}}=0,\sqrt s,p_0,\Delta Y)}{\pi
R_A^2 \tau_0\Delta Y},
\label{epsBJ}
\end{equation}
where the initial time $\tau_0=1/p_0$ and the spatial rapidity
interval in the longitudinal volume element in the denominator has
been taken to be $\Delta Y$. As seen in Fig. \ref{dydepCTQ}, the
initial transverse energy $\bar E_T^{AA}=T_{AA}\sigma\langle
E_T\rangle$ behaves in LO essentially linearly in $\Delta Y$ near the
central rapidities and the $\Delta Y$ dependence is thus cancelled out
in the initial energy density of Eq. (\ref{epsBJ}). Based on the NLO
computation of the cross sections of observable jets \cite{EKS90}, it
could be expected that $\Delta Y$ plays the role of the jet cone $R$
and the NLO results would therefore exhibit a different behaviour
($\sim R^2$ at large $R$) in $\Delta Y$ than the LO results.  From the
point of view of the local energy densities obtained through
Eq. (\ref{epsBJ}) this would be disastrous as $\epsilon_i$ would badly
depend on the $\Delta Y$ chosen. As clearly seen in the figure, this
is now {\em not} the case but the NLO results behave nicely in a
similar way as the LO results do.  This implies that the initial
energy densities are not sensitive to the detailed size of the
rapidity interval.  One should also pay attention to the difference
between the calculation of $E_T$ of ordinary jets and the calculation
of $E_T$ in $\Delta Y$ from minijets considered in here: the jet
calculation introduces an additional dependence on a resolution
parameter, the jet cone radius $R$, which is simply absent in our
considerations.

\subsection{Nuclear effects}

It is a well known fact that the distribution of partons in a free
proton are different from those of a proton bound to a nucleus,
$f_{i/A}(x,\mu^2)= R_i^A(x,\mu^2)f_i(x,\mu^2)$ with $R_i^A\neq 1$.
Regarding the behaviour of $\sigma\langle E_T\rangle$, the last point
of study here is the effect of an inclusion of nuclear effects
$R_i^A(x,Q^2)$ in the parton distributions. This is done by using the
EKS98 parametrization \cite{EKS98} which is based on a DGLAP analysis
of the nuclear parton distributions, with constraints from the
existing data from deep inelastic $lA$ scattering and Drell-Yan
production in $pA$ and also from momentum and baryon number
conservation. There are definite sources of uncertainties we
implicitly accept when using EKS98: First, the nuclear modifications
have been shown to be independent of the PDF set chosen within a level
of a few per cent error \cite{EKS98}. Second, the EKS98 is based on a
LO DGLAP evolution equations only but we apply it to the NLO PDFs as
well. Finally, the nuclear gluon distributions themselves are still
largely unknown at $x\lsim0.01$, a region which is important for the
few-GeV minijets at the LHC. For RHIC, the situation is better since
for $0.01\lsim x\lsim 0.2$ some constraints can be obtained from the
$Q^2$ dependence of the ratio of the structure functions $F_2^{\rm
Sn}/F_2^{\rm C}$ measured by the New Muon collaboration in deeply
inelastic lepton-nucleus scatterings \cite{NMC}.

\begin{figure}[hbt]
\vspace{0.0cm}
\begin{minipage}[b]{0.5\linewidth}
	\centering \includegraphics[width=8cm,trim=100 100 0
	0]{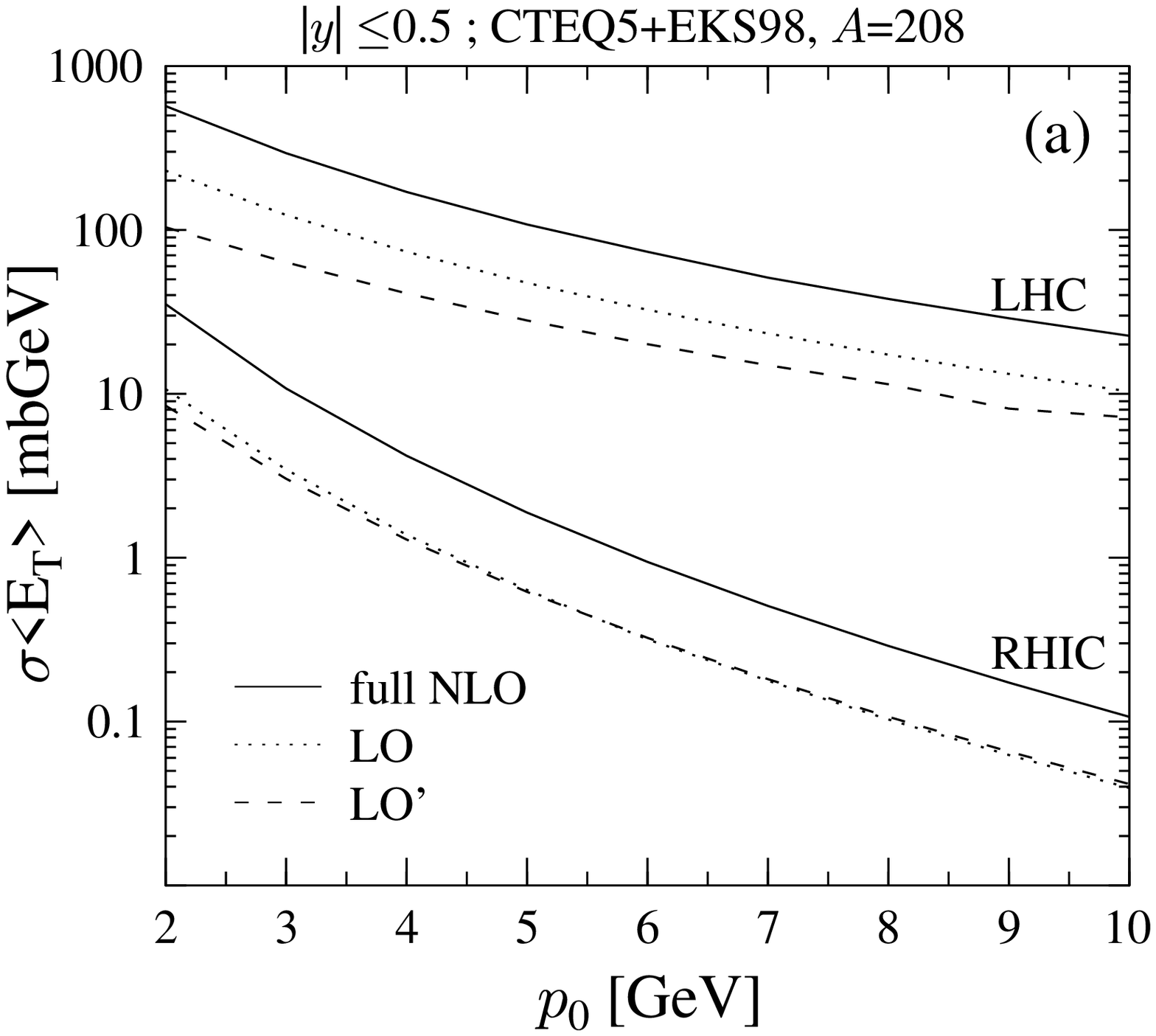} 
\end{minipage}
\hspace{1cm}
\begin{minipage}[b]{0.5\linewidth}
	\centering \includegraphics[width=8cm,trim=100 100 0
	0]{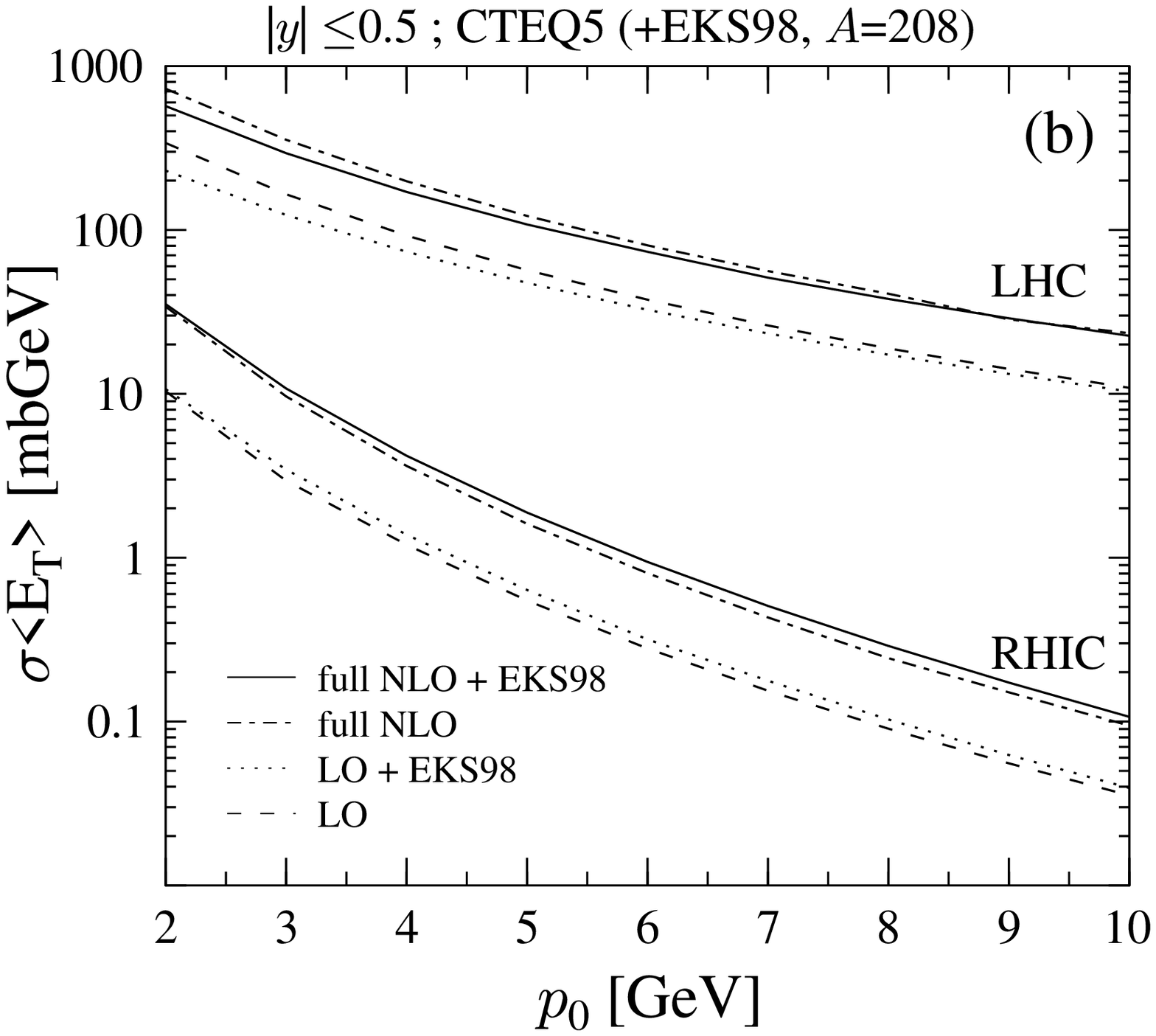} 

\end{minipage}
\caption{(a) \small The $\sigma\langle E_T \rangle$ as in
	fig. \ref{p0dep}(a) but also nuclear shadowing has been taken
	into account via EKS98 parametrization \cite{EKS98}. The
	computation has been carried our for a nucleus $A$=208 for both
	energies.  (b) Comparison of the $\sigma\langle E_T \rangle$
	with nuclear shadowing (solid and dotted lines) and without
	shadowing (dot-dashed and dashed lines).  The curves are shown
	for RHIC (lower four curves) and LHC (upper four curves).  Of
	the four curves the upper pair is the full NLO result and the
	lower pair is the truly LO result.}
\label{shadowkuva}
\end{figure}

The validity region of the EKS98 does not quite extend down to $\mu=1$
GeV, so we limit the study of nuclear effects in the NLO study of
$\sigma\langle E_T\rangle$ to $p_0$ between 2 and 10 GeV. The results
can be found in Fig. \ref{shadowkuva} below. In the first panel, the
nuclear effects are included in all curves, the labelling of which is
the same as in previous figures. The corresponding $K$-factors are
collected in Table \ref{Kshadtable}.  Comparison of the numbers with those
in parentheses in Table 2 indicates that the effects of the nuclear
modifications to the $K$-factors can be neglected at RHIC, whereas for
the LHC there is a slight increase towards the region $p_0\sim 2$
GeV. The reason for this can be understood based on the second panel,
where a comparison between the results with and without the nuclear
effects is shown. For RHIC near $p_0\sim 2$ GeV, one is in the region
where the nuclear effects of gluons stay very small as the cross-over
region between shadowing ($R_g^A<1$) and anti-shadowing ($R_g^A<1$) is
probed (For $2\rightarrow2$, typically $x\sim 2p_0/\sqrt s$). Towards
$p_0\sim 1$ GeV one would move into the shadowing region but the
nuclear effects remain quite small in any case. At the LHC, the shadowing 
effects are larger, about 30\% for the LO and 20\% for the NLO results.
The difference is due to the new kinematical region included: 
in the region $E_T<p_0$, where typically $y_1>0.5$  and $y_2<0.5$
and $p_3$ is small, the values of $x_1$ and $x_2$ are slightly larger than 
the typical $x_{1,2}$ in LO. The net loss in the gluonic luminosity
due to shadowing is then altered less than in LO.

\begin{table}[!hbt]
\begin{center}
\begin{tabular}{|l|c|c||c|c|}
\hline $p_0$ & $K_{RHIC}^{\prime}$ & $K_{RHIC}$ &
$K_{LHC}^{\prime}$ & $K_{LHC}$ \\ 
\hline 
2  & 4.2  & 3.3  & 5.5  & 2.5  \\ 
4  & 3.2  & 3.0  & 4.2  & 2.3  \\
6  & 2.9  & 3.0  & 3.7  & 2.3  \\
8  & 2.7  & 2.7  & 3.3  & 2.2  \\
10 & 2.6  & 2.7  & 3.1  & 2.2  \\
\hline
\end{tabular}
\end{center}
\caption{\small Values of the two $K$-factors defined in
Eq. (\ref{Ks}) for different values of $p_0$. The parton distribution
functions are those of the CTEQ5-set and the nuclear shadowing is
included through the EKS98 parametrization.}
\label{Kshadtable}
\end{table}

\section{From $p_0$ to $p_\rmi{sat}$?}

As described in the introduction, the goal of the present paper is to
study the main uncertainties in the collinearly factorized minijet
formalism to the initial production of transverse energy in
ultrarelativistic nuclear collisions. 
Now that we have control over the NLO corrections in the quantity
$\sigma\langle E_T\rangle$ and have also verified that they do not
present significantly divergent results at scales of few GeV, we may,
as an application, turn to a determination of the actual value of the
scale $p_0$.  As emphasized already in the introduction, this requires
additional phenomenology. Here we will work in the framework of
Ref. \cite{EKRT99}.

At certain scale $p_0=p_\rmi{sat}(\sqrt s,A)$ the system of gluons
produced into the central unit of rapidity will reach a density such
that the effective total area of these gluons $N_{AA}(\sqrt s, p_{\rm
sat})\times \pi/p_\rmi{sat}^2$ will be equal to the transverse area
$\pi R_A^2$ in central collisions.  Performing the computation at a
scale $p_0=p_\rmi{sat}$ will effectively estimate the contributions of
all scales in the sense that gluons with $p_T\ll p_\rmi{sat}$ would
contribute negligibly to the total $E_T$. Hence, if $p_\rmi{sat}$
falls into the perturbatively computable region, above $\sim$ 1 GeV,
the initial energy densities can be obtained from the perturbative
computation alone. For nuclei with $A\sim 200$ and collision energies
of the order of 100 GeV this has been shown to be the case
\cite{EKRT99}.

\begin{figure}[hbt]
\vspace{-0.0cm}
\begin{center}
\includegraphics[width=8cm]{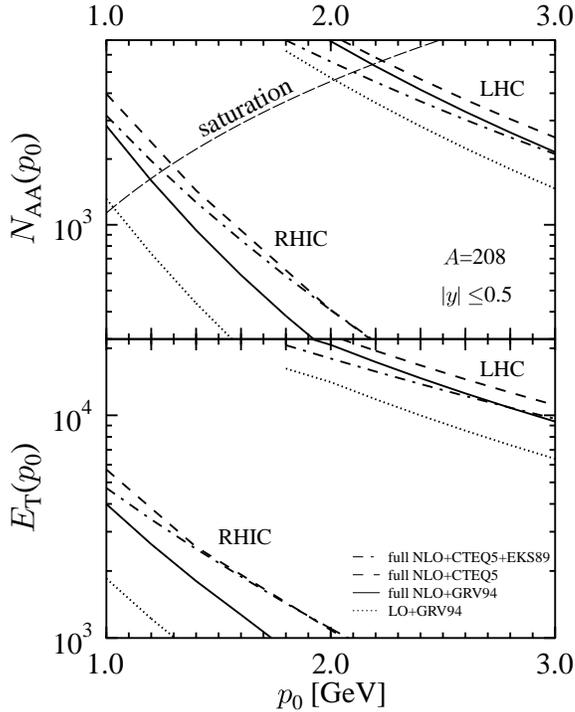}
\caption{\small Upper panel shows the determination of the scale
$p_{\rmi{sat}}$. The saturation curve, $N_{AA}(p_0)=p_0^2R_A^2$ is
shown by the dashed line labeled `saturation'.  2 sets of four curves
are shown. The upper four curves are for LHC energies and the lower
ones are for RHIC. The solid lines are the LO results with GRV94
parton distributions multiplied by $K(p_0)$ from the $\sigma\langle
E_T\rangle$ analysis, and the dotted ones are the LO results with
GRV94 without the $K$-factor. The dashed lines stand for the LO
results with CTEQ5 and the dot-dashed lines for the LO results with
CTEQ5+EKS98, multiplied with respective $K(p_0)$ from the
$\sigma\langle E_T\rangle$ calculation with the same PDFs. The value
of $\psat$ can be read off in the intersection points of the
`saturation'-curve with the other curves. The corresponding values of
$E_T(\psat)$ can be read off from the curves at the lower panel.}
\label{satfig}
\end{center}
\end{figure}

To demonstrate the effects of the NLO terms and shadowing corrections,
a determination of the scale $p_{\rmi{sat}}$ is illustrated in the
upper panel of Fig. \ref{satfig}, showing the number of minijets
in a central $AA$ collision of $A=208$, $N_{AA}(b=0,p_0,\sqrt s,
\Delta Y)=T_{AA}(0)\times 2\sigma_{\rm jet}(p_0,\sqrt s,\Delta Y)$
\cite{EK97} in the rapidity acceptance region $\Delta Y$.  There is,
however, a theoretical caveat in computing the number of minijets, or
rather $2\sigma_{\rm jet}(p_0,\Delta Y)=\sigma\langle n\rangle$, the
first moment of the number distribution of minijets in $\Delta Y$ to
NLO.  A strictly infrared safe definition of a measurement function
for $\sigma\langle n\rangle$ requires an introduction of an additional
resolution scale, such as a screening mass, which defines when two
nearly collinear partons within $\Delta Y$ are to be counted as one
and when as two.  A detailed formulation of this is beyond the scope
of the present paper, so let us simply simulate the NLO terms by taking the
$K$-factors computed for $\sigma\langle E_T\rangle$ and multiply the
LO results for $N_{AA}$ by them. The value of $p_0$ can then be
determined by the intersection points of $N_{AA}(p_0)$ with the saturation
curve $N_{AA}(p_0)=p_0^2 R_A^2$. In the lower panel we show the results for
the initial transverse energy $E_T^{AA}=T_{AA}\sigma\langle
E_T\rangle$ with and without shadowing. The initial $E_T$ can then 
be read off from the different curves at each saturation scale 
$p_0=p_{\rm sat}$.

It is interesting to note, as pointed out already in \cite{EKRT99},
that at the saturation scale $p_0=p_{\rm sat}$ the computed
$E_T^{AA}/N_{AA}$ is very close to $E/N=\epsilon_i/n_i=2.7 T_i$ of an
ideal gas of massless bosons, where the temperature $T$ is obtained
from $\epsilon_i = aT^4$ with $a=16\pi^2/30$.  From the point of view
of energy per particle, a very rapid thermalization seems thus
plausible. Recent studies \cite{MUELLER1,MUELLER2,JEFF} indicate that
this may indeed be the case, although early thermalization is still a
subject of debate \cite{DG00}. In any case, it is worth emphasizing
that in our approach saturation and apparent thermalisation take place
simultaneously.

In fact, the early thermalisation could be used to get rid of the
worry that the saturation approach involves a non-infrared safe
quantity: one could first compute the thermalised initial parton
number $N_i(p_0)$ as a function of $p_0$ from the infrared safe
initial $E_T$ as $E_T^{AA}(p_0) \rightarrow \epsilon_i(p_0)\rightarrow
T_i(p_0)\rightarrow s_i(p_0)\rightarrow N_i(p_0)$, and then use the
obtained thermal $N_i(p_0)$ in the saturation criterion to determine $p_{\rm
sat}$.

If the system is thermal at $p_{\rm sat}$ and if its further evolution
is isentropic, the initial number of particles reflects the final
multiplicity per unit rapidity quite closely, $N_f\approx 0.9
N_{AA}(p_{\rm sat},\sqrt s)$ \cite{EKRT99}. Figure \ref{satfig} can
therefore be directly used to estimate the effects from the NLO
contributions, parton distribution functions and shadowing to
$N_f$. If the NLO corrections are taken into account via the
$K$-factor, then from the saturation criterion $N_{AA}=\psat^2
R_{A}^2$ it follows, assuming that $N_{AA}\sim K/\psat^n$, that
$N_{AA}\sim K^{2/(n+2)}$. Furthermore taking $E_T\sim \psat N_{AA}$
one finds that $E_T\sim K^{3/(n+2)}$. From the Fig. \ref{satfig} we
see that $n=3.0\dots 3.3$ at RHIC and $n=2.4\dots 2.7$ at LHC,
depending whether shadowing is included or not.  Let us take, say,
$K\approx 2$ at RHIC and $K\approx 1.5$ at LHC. Then, using the
smallest values for $n$ quoted above, one finds that at RHIC the
increase with respect to the LO results in $N$ and $E_T$ is 30\% and
50\% respectively. At LHC the increase is 20\% $N$ and 40\% in $E_T$.
This also demonstrates the importance of the inclusion of the NLO
contributions

It is more difficult to give the theoretical uncertainty
in the saturation criterion itself, due to the possible
nonperturbative elements involved. The data coming now for the charged
particle multiplicities from RHIC can in principle be used to
constrain the remaining uncertainty in the saturation criterion. Once
the scale $p_0=p_{\rm sat}$ is nailed down for one collision energy,
predictions of the final state $E_T$ within the hydrodynamical
approach can be computed with the initial energy densities obtainable
based on the infrared safe NLO computation of $E_T^{AA}$.  As
mentioned in the introduction, a large decrease in $E_T$ is expected
due to the work done by pressure. So far only longitudinally expanding
QGP has been studied with minijet initial conditions \cite{EKRT99},
the results from a transversally expanding hydrodynamic system will be
reported soon \cite{ERRT}.

\section{Conclusions}

We have presented in detail the theoretical basis, implementation and
numerical results of a rigorous perturbative QCD calculation of the
amount of transverse energy produced from minijets in $AA$ or $pp$
collisions into a rapidity acceptance region.  The formulation is
based on collinear factorization.  In particular, the quantity
$\sigma\langle E_T\rangle$, the first moment of the minijet $E_T$
distribution, is computed and analysed.  

Theoretically, the key point is the formulation of an infrared safe
measurement function which gives the minijet $E_T$ distribution in
$\Delta Y$ and includes the definition of perturbative collisions. The
singularities arising in the NLO terms are cancelled by applying the
subtraction method \cite{EKS89,KS92}. The dependence of the results on
the parton distribution functions, on the scale choice, on the width
of the acceptance window and on nuclear effects in parton
distributions have been studied in detail. Also, as an application, we
have considered a way to fix the minimum transverse momentum scale
$p_0$ in the saturation model \cite{EKRT99}, and the uncertainties from
this procedure to the initial conditions of the QGP in $AA$ collisions at RHIC
and LHC and also to the charged particle multiplicities in the final
state.

The analysis of the full NLO results leads us to conclude that the NLO
corrections are large but well behaved in the sense that their
relative magnitude as compared to the LO results is not diverging as
subprocesses with smaller and smaller amount of transverse energy
exchanged were taken into account. Furthermore, the contributions from
two different regions $E_T\ge p_0$ and from $E_T\le p_0$ were
studied. The latter one contributes only in NLO, through
configurations where one small-$p_T$ parton enters $\Delta Y$ and the
other two partons with larger transverse momenta fall outside the
acceptance window. When the contribution from the new domain $E_T\le
p_0$ is excluded, the NLO corrections are observed to reduce by roughly
a factor of two, and the stability of the results against the LO to
increase. If the new domain is included, the comparison of the full
NLO results against the LO numbers is not strictly one-to-one, since
the new kinematical region is not present in LO at all. A consistent
comparison in this case can therefore only be performed between NNLO
and NLO contributions, as then contributions from all values of $p_T$
would be present. Our conjecture is that in that case suppression of
NNLO terms will be observed.

Another important and beforehand not at all obvious result is that the
initially produced $E_T$ in $AA$ collisions depends, contrary to what one
might expect based on the jet cone dependence of high-$p_T$ jets,
practically linearly on the chosen rapidity window $\Delta Y$. This
has the important consequence that the local energy density does not
depend on $\Delta Y$, and a possible source of ambiguity in estimation
of the initial energy densities for the QGP in high energy $AA$
collisions is removed.

Finally, in connection to other models, in particular to the
McLerran-Venugopalan model for gluon production through classical
fields \cite{McLV94}, the inclusion of the new kinematical region in our
computation should bring these two approaches closer to each other.

\vspace{1cm}
\noindent{\bf Acknowledgements.} We thank V. Ruuskanen for discussions
and K. Kajantie for discussions and for collaboration in the early
stages of the project. We thank the Center of Scientific Computing
(CSC, Finland) for providing computational resources and
P. Raiskinm\"aki for guidance in parallelisation of our code. The
financial support from the Academy of Finland as well as from the
Waldemar von Frenckell foundation is gratefully acknowledged.

\end{document}